\documentclass[prd,showpacs,preprint]{revtex4}
\usepackage{amssymb}
\usepackage{amsmath}
\usepackage{graphicx}

\begin{document}

\title{Same-sign top pair production in an extra-dimension model of flavor at the CERN Large Hadron Collider}
\author{Jun Gao}
\email{gaojun49@pku.edu.cn}
\author{Chong Sheng Li}
\email{csli@pku.edu.cn}
\author{Xiangdong Gao}
\email{gaoxiangdong@pku.edu.cn}
\author{Zhao Li}
\email{zhli.phy@pku.edu.cn} \affiliation{Department of Physics and
State Key Laboratory of Nuclear Physics and Technology, Peking
University, Beijing 100871, China}

\date{\today}


\begin{abstract}
We study the same-sign top pair production mediated by the first
Kluza-Klein (KK) excitation of the gluon in the Randall-Sundrum (RS)
model with flavor violation at the Large Hadron Collider (LHC), in
which the nonuniversal couplings between fermions and KK gauge
bosons will lead to observable tree level flavor-changing neutral
current (FCNC) effects. We find that the same-sign top quarks
produced in our case have property of high energy and high
transverse momentum, and lead to an observable signal in the
same-sign dilepton channel even when the mass of the KK gluon reach
up to 3 TeV. We further investigate the potential of the LHC to
probe the flavor violating parameters and find that the LHC can
probe their values down to 0.06.
\end{abstract}

\maketitle

\section{introduction}
Now search for extra dimensions has been one of the major objects at
the LHC, since its physical effects can appear at the TeV energy
scale. The idea of extra dimensions was revived in the 1990's
~\cite{Antoniadis:1990ew,ArkaniHamed:1998rs,Randall:1999ee}, which
brings new solutions to the gauge hierarchy problem and also can be
used to resolve the fermion mass hierarchy problem. The RS
model~\cite{Randall:1999ee} with a warped geometry in five
dimensions is one of the most important cases. In the RS model, the
single extra dimension is compactified on a $\rm S^1/Z_2$ orbifold
with a radius $r$, which is not too large compared with the Planck
length. Two 3-branes, the Planck brane and the TeV brane, are
located at the orbifold fixed points $\phi=0,\pi$, respectively, and
the spacetime between the two 3-branes is simply a slice of a
five-dimensional anti-de Sitter ($\rm AdS_5$) geometry. The
five-dimensional warped metric is given by
\begin{equation}
ds^2=e^{-2kr|\phi|}\eta_{\mu \nu}dx^{\mu}dx^{\nu}-r^2d\phi^2,
\end{equation}
where $\phi$ is the five-dimensional coordinate, and $k \sim M_P$ is
the curvature scale. By requiring $kr\sim 12$, one can suppress the
Planck scale to $ M_P e^{-k\pi r}\sim O({\rm TeV})$ on the TeV
brane, and then solve the gauge hierarchy problem.

The original RS model has also been generalized to allow the
standard model (SM) fields to reside in the bulk
~\cite{Goldberger:1999wh,Chang:1999nh,Grossman:1999ra,
Gherghetta:2000qt,Huber:2000ie,Davoudiasl:2000wi,Hewett:2002fe},
which can generate the fermion mass hierarchy by exponential warped
factors. For each fermion flavor $i$, we have two five-dimensional
Dirac fermions $\Psi_{i,L}(x,y)$, and $\Psi_{i,R}(x,y)$,
($y=r\phi$), while for simplicity the Higgs field will be localized
on the TeV brane. Thus, the Yukawa coupling of bulk fermions can be
expressed as~\cite{Gherghetta:2000qt}
\begin{eqnarray}
\int d^4x dy\sqrt {-g} \lambda_{ij}^{(5)}\left (H(x) \bar
{\Psi}_{i,L}(x,y)\Psi_{j,R}(x,y)+{\rm h.c.} \right)\delta(y-\pi
r)\nonumber\\
\equiv \int
d^4x\lambda_{ij}\left(H(x)\bar{\Psi}_{iL}^{(0)}(x)\Psi_{jR}^{(0)}(x)+h.c.+...
\right),
\end{eqnarray}
where $\lambda_{ij}^{(5)}$ are the five-dimensional Yukawa couplings
and $\lambda_{ij}$ are the effective four-dimensional Yukawa
couplings between the Higgs fields and the SM fermions,
$\Psi_{iL}^{(0)}$ and $\Psi_{jR}^{(0)}$, which correspond to the
zero KK modes. If each fermion field has a 5D bulk mass, which can
be parameterized as $M_{f(i,j)(L,R)}\equiv kc_{(i,j)(L,R)}$, then we
can obtain~\cite{Gherghetta:2000qt}
\begin{equation}
\lambda_{ij}=\frac{\lambda_{ij}^{(5)}k}{N_{iL}N_{jR}}e^{(1-c_{iL}-c_{jR})\pi
kr},
\end{equation}
with
\begin{equation}
\frac{1}{N_{iL}^2}\equiv \frac{1/2-c_{iL}}{e^{(1-2c_{iL})\pi kr}-1},
\end{equation}
and similarly for $N_{jR}$. For light fermions, by setting
$c_{iL},c_{jR}\gtrsim 1/2$, which means the light fermions are
localized towards the Planck brane, one can generate
exponentially-small Yukawa couplings. On the other hand, the third
generation quarks can be close to the TeV brane and have $O(1)$
Yukawa couplings by requiring $c_{iL},c_{jR}<1/2$.

Since the light fermions live towards the Planck brane, their
couplings to KK gauge bosons will be small and universal. But for
the right-handed top quark and the left-handed doublet $Q^3$, they
could have rather strong couplings to the KK gauge
bosons~\cite{Gherghetta:2000qt,Aquino:2006vp}. These nonuniversal
couplings will lead to observable FCNC effects at tree level, which
have been extensively studied in
Refs~\cite{Aquino:2006vp,Davoudiasl:2001uj}.

In the SM, the same-sign top pair production rate is highly
suppressed due to the GIM mechanism, but can be enhanced to an
observable level in some new physics models, for example,
supersymmetric standard model~\cite{Hou:1995qh}, topcolor-assisted
technicolor model~\cite{Cao:2004wd}, and maximal flavor violation
model~\cite{BarShalom:2008fq}. The same-sign top pair can also be
produced with the model-independent FCNC couplings, $gtq$, $Ztq$,
$\gamma tq$, $Htq$ and $Z'tq$~\cite{Gouz:1998rk}. All these
processes can provide very clean signals due to the small SM
backgrounds for the same-sign dilepton channel. In this paper we
study the same-sign top pair production mediated by the first KK
excitation of the gluon, $G^{(1)}$, at the LHC and use this process
to directly probe the FCNC couplings between the up-type quarks and
the first KK gluon. And we show the production rate can be greatly
enhanced due to the large FCNC couplings of the up-type quarks and
the KK gluon.

The arrangement of this paper is as follows. In Sec.~II, we give the
relevant FCNC couplings and parameters. In Sec.~III, we show the
cuts for signal selection and numerical results. Sec.~IV contains a
brief conclusion.

\section{FCNC couplings}
In general, the top FCNC couplings in our case can be written as
\begin{eqnarray}
\mathcal{L}_{FCNC}&=&g_{t_R}U_R^{tt*}U_R^{tc}(\bar{t}_R
T^a\gamma^{\mu}c_R)G_{\mu}^{(1)a}+
g_{t_R}U_R^{tt*}U_R^{tu}(\bar{t}_R T^a\gamma^{\mu}u_R)G_{\mu}^{(1)a}\nonumber\\
&+&g_{t_L}U_L^{tt*}U_L^{tc}(\bar{t}_L
T^a\gamma^{\mu}c_L)G_{\mu}^{(1)a}+
g_{t_L}U_L^{tt*}U_L^{tu}(\bar{t}_L
T^a\gamma^{\mu}u_L)G_{\mu}^{(1)a}+h.c.,
\end{eqnarray}
where $g_{t_L}$, $g_{t_R}$ are the coupling constants between top
quark and $G^{(1)}$, and $U_L,U_R$ are the left-handed and
right-handed rotation matrices that transform the up-type quarks
from the weak eigenstate basis to the mass eigenstate basis. Other
contributions from the first two generations can be neglected due to
the smallness of the coupling constants. As pointed out in
Refs.~\cite{Agashe:2006at,Cacciapaglia:2006gp,Lillie:2007ve,Cacciapaglia:2006mz},
a KK gluon with mass $M_{G^{(1)}}$ as low as 1 TeV is still
possible, so we will consider $M_{G^{(1)}}$ values from 1 TeV to 3
TeV below. Within this mass range, the authors of
Ref.~\cite{Aquino:2006vp} discuss some interesting values of $c_L^3$
and $c_R^t$ which can generate the correct quark masses, mixings and
also satisfy all the electroweak precision constraints, and give a
possible range of the coupling constants, $g_{t_L}=[1.0,2.8]g_s$,
$g_{t_R}=[1.5,5]g_s$, where $g_s$ is the usual 4D strong interaction
coupling constant.

The rotation matrixes $U_L$ and $U_R$ depend on the 5D Yukawa
couplings and the bulk masses of the up-type quarks. In principle,
we have little knowledge on the matrix elements, and the choice of
their values has a great freedom. However, in order to generate the
correct Cabibbo-Kobayashi-Maskawa (CKM) matrix, we have
$V_{CKM}=U_L^\dagger D_L$. By assuming $U_L \sim \sqrt{V_{CKM}}$,
$U_L^{tc}$ and $U_L^{tu}$ are very small~\cite{Aquino:2006vp}, so we
will neglect the contributions from the left-handed FCNC couplings
in our following calculations. On the other hand, there is no direct
constraint on the elements of $U_R$, so we treat $U_R^{tt}$,
$U_R^{tc}$ and $U_R^{tu}$ as free real parameters as in
Ref.~\cite{Aquino:2006vp}, and investigate the LHC reach of the
parameter region.

\begin{figure}[!ht]
\includegraphics[width=0.4\textwidth]{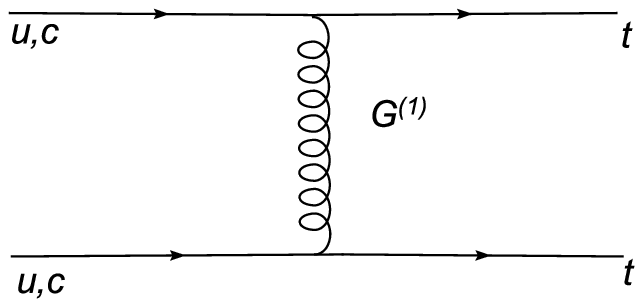}
\caption[]{Feynman diagram for the same-sign top pair production
with FCNC couplings. There is also a corresponding u-channel
diagram.} \label{f1}
\end{figure}

The relevant Feynman diagrams for same-sign top pair production are
shown in Fig.~\ref{f1}. At the LHC, the main contribution to the
same-sign top pair production arises from subprocess $uu\rightarrow
tt$ due to the high parton luminosity of the $u$ quark. However, in
order to probe $U_R^{tc}$, we also include contributions from the
subprocesses $uc,cu,cc\rightarrow tt$. We define two flavor
violating parameters $\varepsilon _u\equiv |U_R^{tt}U_R^{tu}|$, and
$\varepsilon _c\equiv |U_R^{tt}U_R^{tc}|$, which must satisfy
$\varepsilon _c^2 + \varepsilon _u^2\leq0.5^2$ due to the unitarity
of $U_R$. Note that our study is almost model-independent and
depends only on $\varepsilon_u$, $\varepsilon_c$, $g_{t_R}$ and
$M_{G^{(1)}}$. In our numerical calculations, the CTEQ6L1 PDF
set~\cite{Pumplin:2002vw} is used, and renormalization and
factorization scales are set to the top quark mass.

\begin{figure}[!ht]
\includegraphics[width=0.9\textwidth]{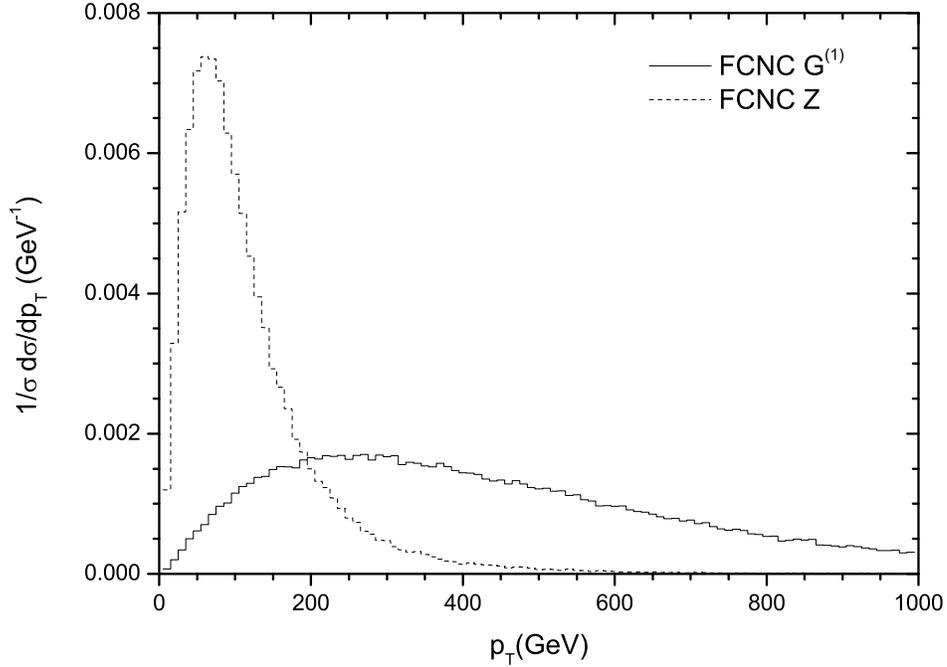}
\caption[]{Normalized top quark transverse momentum distribution in
the same-sign top pair production with FCNC couplings mediated by KK
gluon and Z boson at the LHC, assuming $\varepsilon_u=0.5$,
$\varepsilon_c=0$ and $M_{G^{(1)}}=1{\rm TeV}$.} \label{f2}
\end{figure}

In Fig.~\ref{f2} we show the transverse momentum distributions of
the top quark in the same-sign top pair production at the LHC with
the FCNC couplings mediated by Z boson and KK gluon, respectively,
assuming $M_{G^{(1)}}=1\rm TeV$. Due to the large mass difference,
the top quarks are typically produced with much higher energy and
transverse momentum in KK gluon exchange than in Z boson exchange.
These properties can be further used to help us suppress the SM
backgrounds. The total cross sections of the same-sign top pair
production mediated by KK gluon at the LHC are given in
Table~\ref{t1}, which shows the total cross section can reach as
high as 16 pb.

\begin{table}
\begin{center}
\begin{tabular}{|c|c|c|c|}
\hline  $M_{G^{(1)}}$ &~~~1TeV~~~&~~~2TeV~~~&~~~3TeV~~~
\\
\hline $\varepsilon_u=0.5,\varepsilon_c=0$   &16.4&1.71&0.41
\\
\hline $\varepsilon_u=\varepsilon_c=0.35$    &4.75&0.48&0.11
\\
\hline
\end{tabular}
\end{center}
\caption{The total cross sections (in pb) for $tt$ production at the
LHC under different values of the KK gluon mass and flavor violating
parameters, assuming $g_{t_R} = 3g_s$.} \label{t1}
\end{table}

\section{signal and backgrounds}
In order to distinguish our process from the SM $t\bar t$
production, we only consider the leptonic decays of the top quarks,
$pp\rightarrow tt \rightarrow bbl^+l^+\nu_l \nu_l$, where $l=e$ or
$\mu$. The resulting signal consists of two $b$ jets, two positive
charged leptons, missing transverse energy and possible light jets
from showering. We will combine two channels, one of which contains
exactly one $b$-tagged jet in the final state, and the other
contains two $b$-tagged jets. There are several main SM backgrounds
for our process: (a) $pp\rightarrow W^+t\bar t$, $W^+W^+qq$, (b)
$pp\rightarrow W^+Zqq$, when the $l^-$ from Z decay is undetected.
Other possible backgrounds like $pp\rightarrow W^+W^+W^-$, $ZZqq$,
$W^+b\bar b$, $tW^-(\bar t W^+)$, and $t\bar t$ are very small
according to our calculations, and can be neglected. The above
backgrounds are simulated with MADEVENT~\cite{Maltoni:2002qb}, and
the signal events are generated by COMPHEP 4.4~\cite{Boos:2004kh}.
In our calculations, PYTHIA 6.4~\cite{Sjostrand:2006za} is used to
treat parton showering, hadronization, and PGS4~\cite{conway} is
used for detector simulations, in which b-tagging efficiency has
been taken into account.

We use the following basic acceptance cuts on jets, leptons and
missing transverse momentum,
\begin{eqnarray}
&&p_T(j)>15{\rm GeV},\quad |\eta (j)|<3.0,\quad p_T(l)>10{\rm
GeV},\nonumber\\
&&|\eta (l)|<2.4,\quad \Delta R_{jj}>0.7,\quad
\Delta R_{jl}>0.4,\nonumber\\
&&\Delta R_{ll}>0.4,\quad \not{\! p}_T>20{\rm GeV},
\end{eqnarray}
where $j$ can be either a light jet or a $b$ jet. In order to reduce
the backgrounds in which the charged lepton comes from $b$ decay, we
impose the lepton isolation cuts using PGS4, which are especially
important for suppressing $t\bar t$ backgrounds.

For the one $b$-tagged jet channel, we require the final state
containing exactly one $b$-tagged jet, two positive charged leptons,
and at most two light jets. Besides, we further require
\begin{eqnarray}
&&p_T(j_{max})>20{\rm GeV}, \quad p_T(b)>20{\rm GeV}, \quad p_T(l
_{max})>50 {\rm GeV},\nonumber\\
&&p_T(l_{min})>30 {\rm GeV},\quad H_T>300{\rm GeV},\quad
m(llbj)>400{\rm GeV},
\end{eqnarray}
where we use $j_{max}$ to stand for the leading light jet, $H_T$ is
the sum of transverse energy and $m(llbj)$ is the invariant mass of
all the leptons and jets. As mentioned before, the final state of
our signal comes from two almost back-to-back high $p_T$ top quarks,
so the above cuts can improve the significance greatly. In order to
further reduce the backgrounds, we use the following angular cuts,
\begin{eqnarray}
&&\Delta \phi_{ll}>2.2,\quad \Delta \phi_{bj_{max}}>1.0,
\end{eqnarray}
and require at least one combination of the leptons satisfies
\begin{eqnarray}
&&\Delta R_{lb}+\Delta R_{l'j_{max}}<3.0.
\end{eqnarray}
Some of the backgrounds contain one additional W boson which decays
to two light jets, so we require the light jets invariant mass
should not be within the W mass window:
\begin{eqnarray}
&&m_{jj}<60{\rm GeV}\quad {\rm or}\quad m_{jj}>100{\rm GeV}.
\end{eqnarray}
As for the two $b$-tagged jets channel, we require the final state
containing two $b$-tagged jets, two positive charged leptons, and at
most one light jet. All the additional cuts can be obtained by
replacing the leading light jet in the one $b$-tagged channel cuts
with a $b$ jet.

\begin{table}
\begin{center}
\begin{tabular}{|c|c|c|c|c|c|c|}
\hline Process & signal(1TeV) & signal(2TeV) &signal (3TeV) & $pp
\to W^+t\bar t$ & $pp \to W^+W^+qq$ & $pp \to W^+Zqq$
\\
\hline $\varepsilon_u=0.5,\varepsilon_c=0$   &  28.5  & 3.03&
0.62&0.022&0.003&0.008
\\
\hline $\varepsilon_u=\varepsilon_c=0.35$    &  8.2  & 0.83&
0.17&0.022&0.003&0.008
\\
\hline
\end{tabular}
\end{center}
\caption{Signal and backgrounds cross sections (in fb) after all the
cuts under different values of the KK gluon mass and flavor
violating parameters, assuming $g_{t_R} = 3g_s$.} \label{t2}
\end{table}

\begin{figure}[!ht]
\includegraphics[width=0.7\textwidth]{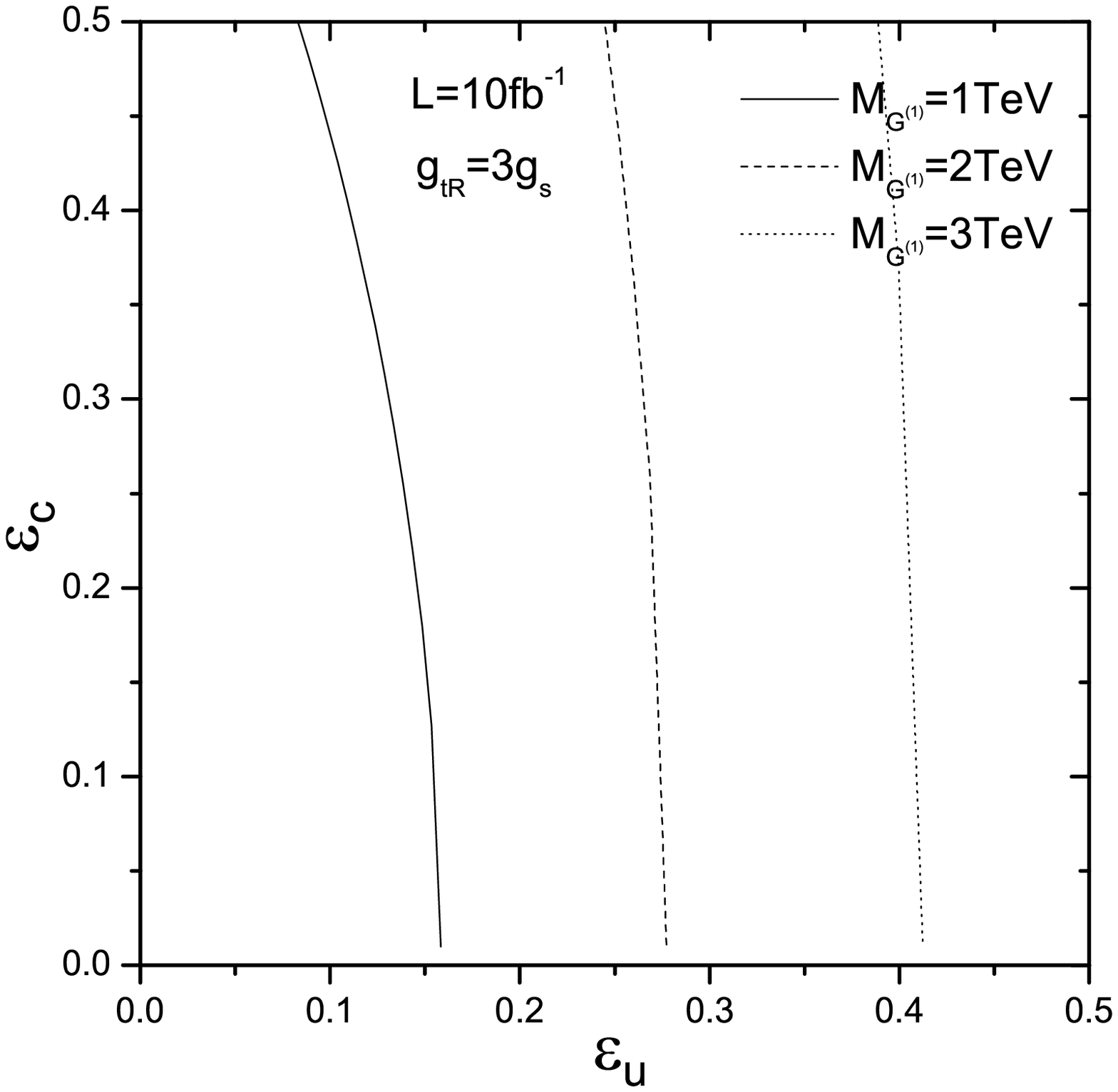}
\includegraphics[width=0.7\textwidth]{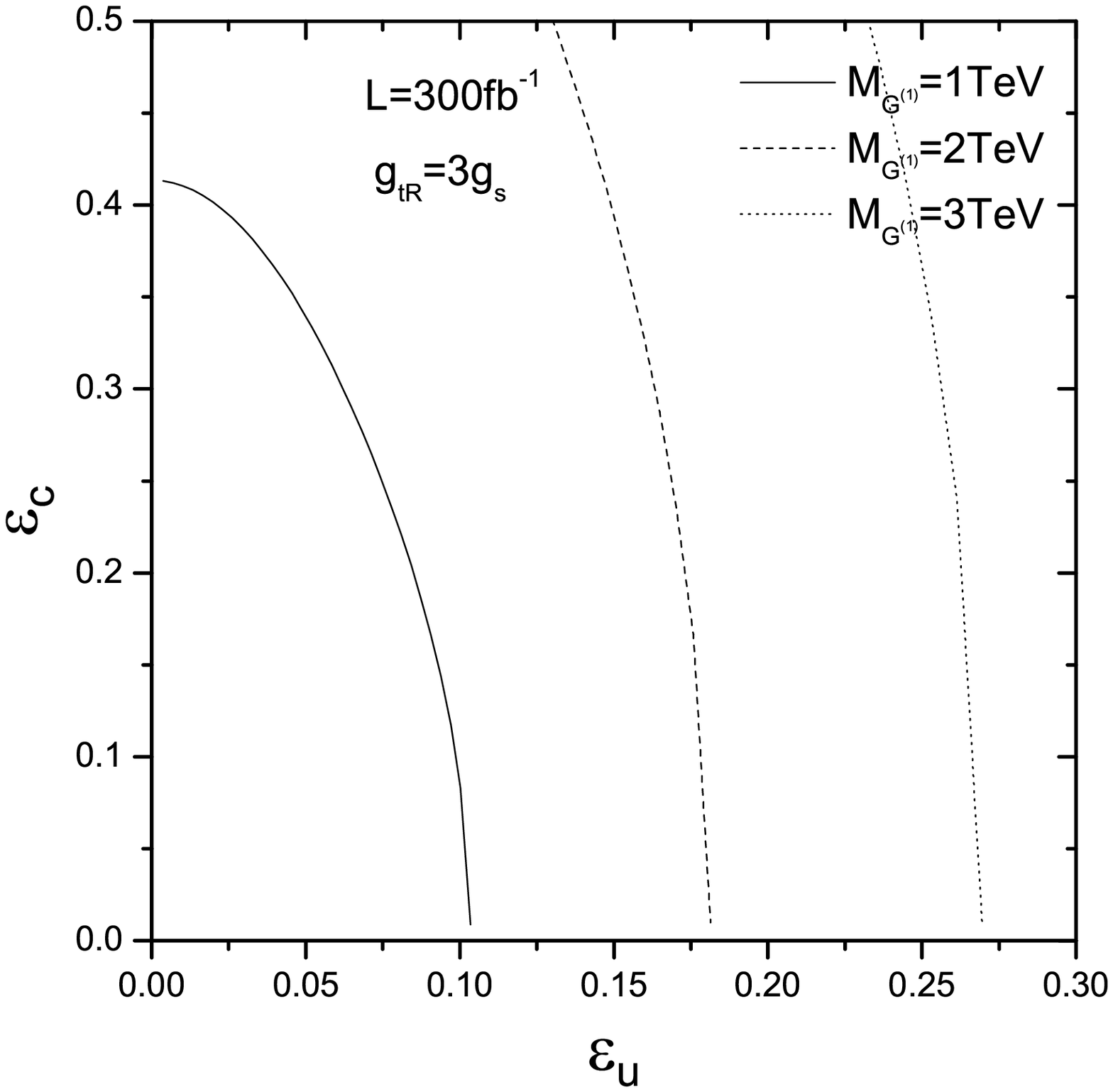}
\caption[]{$5\sigma$ discovery limit of $\varepsilon_u$ and
$\varepsilon_c$ at the LHC for low and high integrated luminosities,
assuming $g_{t_R}=3g_s$.} \label{f3}
\end{figure}

\begin{figure}[!ht]
\includegraphics[width=0.7\textwidth]{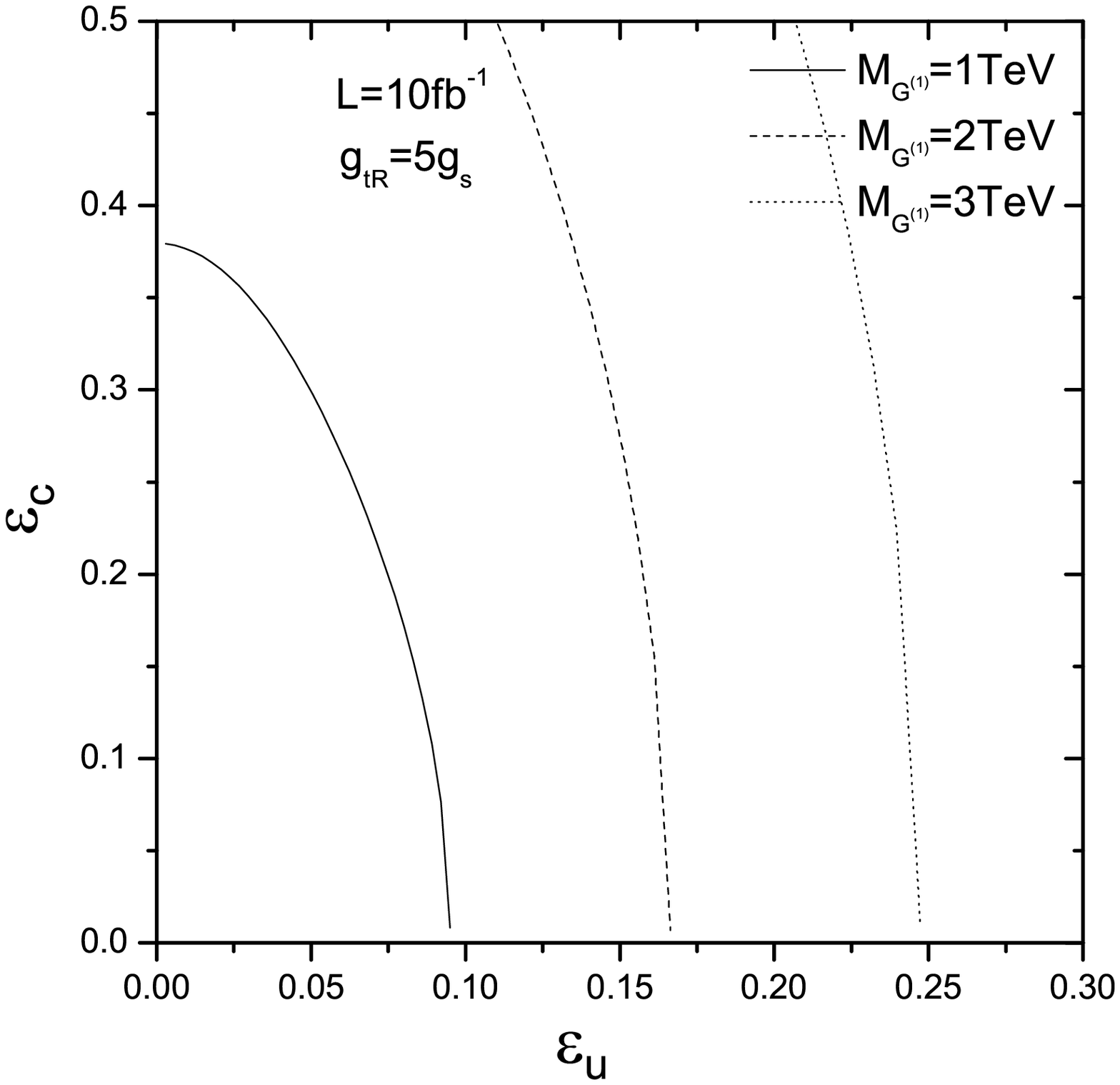}
\includegraphics[width=0.7\textwidth]{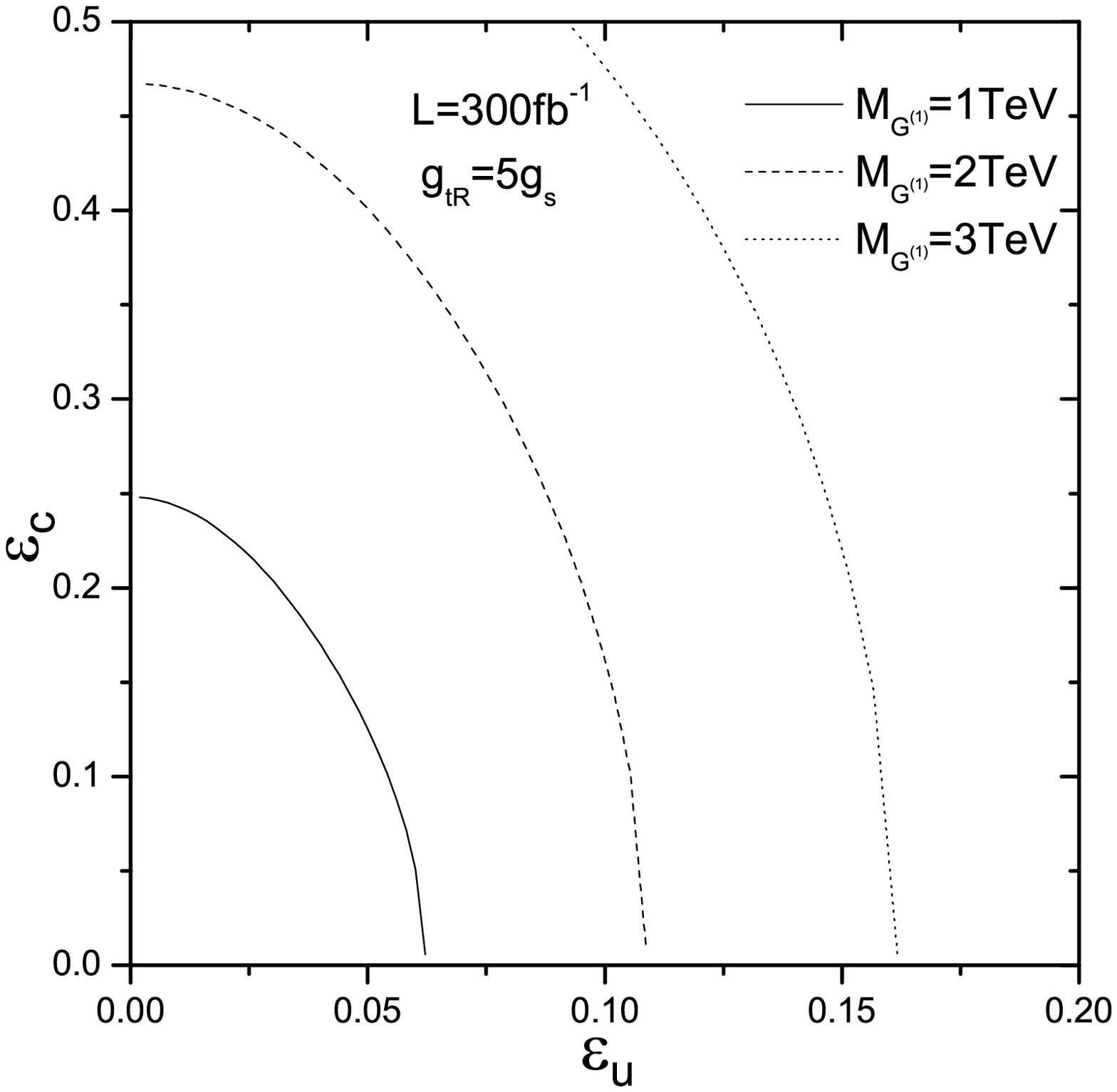}
\caption[]{$5\sigma$ discovery limit of $\varepsilon_u$ and
$\varepsilon_c$ at the LHC for low and high integrated luminosities,
assuming $g_{t_R}=5g_s$.} \label{f4}
\end{figure}

Table~\ref{t2} shows the total cross sections for the signals and
backgrounds with $g_{t_R}=3g_s$ after imposing all the cuts. We can
see that the total backgrounds can be reduced to 0.03 fb, while the
signal can reach 28 fb and 0.6 fb for $M_{G^{(1)}}=1 {\rm TeV}$ and
$3 {\rm TeV}$, respectively. To further investigate the LHC reach we
show in Fig.~\ref{f3} and~\ref{f4} the $5\sigma$ discovery limits of
$\varepsilon_u$ and $\varepsilon_c$ at the LHC, and we see that the
LHC can probe an interesting region of $\varepsilon_u$ and
$\varepsilon_c$ for $M_{G^{(1)}}$ up to $3{\rm TeV}$. Obviously, as
mentioned above, our process is sensitive to $\varepsilon_u$ due to
the large contribution from the subprocess $uu\rightarrow tt$, and
not for $\varepsilon_c$. From Fig.~\ref{f4} we find that the
discovery limit of $\varepsilon_u$ can reach 0.1 for $g_{t_R}=5g_s,
M_{G^{(1)}}=1 {\rm TeV}$ and an integrated luminosity of $10 \rm
fb^{-1}$, while the discovery limit of $\varepsilon_u$ can be as low
as 0.06 for a high integrated luminosity of $300 \rm fb^{-1}$.
Comparing with the results in Ref.~\cite{Aquino:2006vp}, ours are
better for low luminosities and high KK gluon masses. Moreover,
unlike Ref.~\cite{Aquino:2006vp}, all the cuts we imposed are
independent of the KK gluon mass and width, which allow us to probe
the flavor violating parameters for a wide range of $g_{t_R}$ and
$M_{G^{(1)}}$.

\section{conclusions}
In conclusion, we have studied the production of same-sign top pairs
mediated by the KK gluon in the RS model with fermions and gauge
bosons in the bulk at the LHC. For the same-sign dilepton channel,
by imposing suitable cuts the backgrounds can be suppressed to about
0.03~fb, which can lead to a sizable signal/background ratio, about
5$\sim$860. We also investigated the LHC reach of the flavor
violating parameters $\varepsilon_u$ and $\varepsilon_c$, and found
that the LHC can probe an interesting region of the parameters even
when the KK gluon mass is as large as 3 TeV. The discovery limit of
$\varepsilon_u$ can reach 0.1 for $g_{t_R}=5g_s, M_{G^{(1)}}=1 {\rm
TeV}$ and an integrated luminosity of $10 \rm fb^{-1}$. While for a
high integrated luminosity of $300 \rm fb^{-1}$, the discovery limit
of $\varepsilon_u$ can reach as low as 0.06.

\begin{acknowledgments}
This work was supported in part by the National Natural Science
Foundation of China, under Grants No.10721063, No.10575001 and
No.10635030.
\end{acknowledgments}

\newpage
\bibliography{toppair}
\newpage

\end{document}